\documentclass[published]{rrparticle}
\usepackage{amsmath}
\usepackage{multicol}
\usepackage{epsfig}
\usepackage{graphicx}
\usepackage{float}
\usepackage{color}

\title{Bose-Einstein Condensation: Twenty Years After}
\author[1]{V. S. Bagnato}
\author[2]{D. J. Frantzeskakis}
\author[3,4]{P. G. Kevrekidis}
\author[5,*]{B. A. Malomed}
\authorPar[6]{D. Mihalache}
\affil[1]{Instituto de F{\'i}sica de Sao Carlos, Universidade de Sao Paulo,
Caixa Postal 369, \\ 13560-970 Sao Carlos, Sao Paulo, Brazil}
\affil[2]{Department of Physics, University of Athens, Panepistimiopolis,
Zografos, Athens 157 84, Greece}
\affil[3]{Department of Mathematics and Statistics, University of
Massachusetts, \\ Amherst, MA 01003-9305, USA}
\affil[4]{Center for Nonlinear Studies and Theoretical Division, Los Alamos
National Laboratory, \\ Los Alamos, NM 87544, USA}
\affil[5]{Department of Physical Electronics, School of Electrical Engineering,
Faculty of Engineering, \\ Tel Aviv University, Tel Aviv 69978, Israel}
\affil[6]{Horia Hulubei National Institute for Physics and Nuclear Engineering,
P. O. Box MG-6, \\ RO-077125 Magurele, Romania \\ $^*$Corresponding author \email{malomed@post.tau.ac.il}}

\begin{document}
\RRPVolume{67}{2015}
\RRPNumber{1}
\RRPPages{5}{50}
\date{}
\colontitle{Bose-Einstein condensation: Twenty years after}

\maketitle
\begin{abstract}
The aim of this introductory article is two-fold. First, we aim to offer a
general introduction to the theme of Bose-Einstein condensates, and briefly
discuss the evolution of a number of relevant research directions during the
last two decades. Second, we introduce and present the articles that
appear in this Special Volume of \emph{Romanian Reports in Physics}
celebrating the conclusion of the second decade since the experimental
creation of Bose-Einstein condensation in ultracold gases of alkali-metal
atoms.
\end{abstract}

\section{Introduction}

\subsection{Atomic Bose-Einstein condensates}

The Bose-Einstein condensate (BEC) is a macroscopic quantum state of matter,
which was predicted theoretically by Bose and Einstein 90 years ago \cite{1}%
. Atomic BECs were created experimentally in ultracold vapors of $^{87}$Rb
\cite{2}, $^{23}$Na \cite{4} and $^{7}$Li \cite{3} 70 years later. The aim
of this Special Issue is to celebrate the twentieth anniversary of this
remarkable achievement, which was also marked by the Nobel Prize in Physics
for 2001, awarded to E. A. Cornell, W. Ketterle, and C. E. Weiman \cite%
{becnl1}.

In the BEC state, all atoms in the bosonic gas fall (``condense'') into a
single quantum-mechanical ground state. The transition to the BEC occurs if
the atomic density, $n$, and the de Broglie wavelength, $\lambda $,
corresponding to the characteristic velocity of the thermal motion of the
atoms, satisfy the following condition \cite{PethickSmith,PitStrin}:
\begin{equation}
n\lambda ^{3}>2.612,  \label{degen}
\end{equation}
which implies that $\lambda $ is comparable to or larger than the mean
distance between atoms, thus making the gas a macroscopic (degenerate)
quantum state. In the above-mentioned atomic gases, this theoretical
condition is met at temperatures $T$ which are a small fraction of
milli-Kelvin (mK), hence the atomic BECs created in the laboratories are, as
a matter of fact, the coldest objects existing in the universe (the early
BEC-experiments achieving the condensed state around 100 nK). Their creation
became possible after the development of appropriate experimental techniques
needed to reach the necessary ultra-low temperatures (see, e.g., Ref.~%
\cite{laser-cooling}). The required extreme cooling is achieved in two
stages. First, the method of \textit{laser cooling} 
(which was also rewarded with the Nobel Prize in Physics for 1997 \cite{lascool})
is applied to the gas loaded into a magneto-optical trap, which makes it
possible to create a moderately cool state, at temperature $\sim 100$~$\mu$%
K. Next, this state undergoes forced \textit{evaporative cooling}, losing $%
\sim 90\%$ of atoms, and the remaining atomic cloud spontaneously forms the
BEC. In the experiments, the number of atoms in the BEC typically ranges
between $1,500$ and $1,000,000$ (although both smaller and larger numbers
are, in principle, possible), and the size of the domain in which the gas is
trapped is $\sim 100~\mu$m. A characteristic time scale relevant to the
experiments is measured in milliseconds, while the lifetime of the
condensate can be easily raised to several seconds.

The applicability of the laser-cooling method to particular atomic species
depends on the peculiarities of their electron configuration. As a result, this
technique has made it possible to achieve the Bose-Einstein condensation in
vapors of alkaline, alkaline-earth, and lanthanoid metals: $^{7}$Li, $^{23}$%
Na, $^{39}$K, $^{41}$K, $^{85}$Rb, $^{87}$Rb, $^{133}$Cs, $^{52}$Cr, $^{40}$%
Ca, $^{84}$Sr, $^{86}$Sr, $^{88}$Sr, $^{174}$Yb, $^{164}$Dy, and $^{168}$Er.
Perhaps especially interesting, among the more recent developments, is the
creation of BEC in the gases of chromium \cite{Pfau} and dysprosium \cite%
{Lev}, where atoms carry large magnetic moments, which makes it possible to
predict and observe many effects produced by long-range dipole-dipole
interactions \cite{DD-review}. The challenging aim of creating BEC in the
gas of spin-polarized hydrogen atoms has been finally achieved too, with a
specially devised technique which made it possible to cool the gas to $50$ $%
\mathrm{\mu }$K \cite{H}.

The BECs created in the laboratory constitute a prototypical
manifestly quan\-tum-macroscopic state of matter available in the experiments.
In other settings, where low temperatures are crucially important too, macroscopic
quantum effects, such as superconductivity in metals and superfluidity in
liquid helium, are well known too, but they correspond to ``implicit''
quantum states. For instance, a superconducting metallic sample as a whole
is not a macroscopic quantum object. The same pertains to the recently
created out-of-equilibrium BEC of quasi-particles in condensed matter,
namely, exciton-polaritons \cite{67} and magnons \cite{68},
which have drawn a great deal of attention in the past decade (see Refs.
\cite{quasi2,quasi4,quasi5}). Also, a considerable attention was drawn to
the topic of localization of exciton-polaritons in semiconductor
microcavities \cite{quasi6,quasi7,quasi8,quasi9}; for an excellent recent
review focused on several physical phenomena exhibited by exciton-polariton
condensates see Ref. \cite{quasi_review}. The condensation of effectively
massive photons trapped in a microcavity was reported too \cite{photons},
the peculiarity of these settings being the nonconservation of the total
number of the quasi-particles or photons.

Surveys of the broad subject of Bose-Einstein condensation and numerous related areas are provided
by many review articles and books \cite{PethickSmith,PitStrin}, \cite%
{DD-review}, \cite{vsb}-\cite{gauge}.
It is important to mention that this list of surveys on the topic of BEC is,
of course, far from being exhaustive. This is a clear indication of the
impact of this research theme to almost all branches of contemporary physics.

The goal of the present article is to offer a broad picture of
some of the past and currently active research areas in the realm of BEC
(admittedly biased towards the particular research interests of the
authors) and to overview the scientific literature, akin to a Resource Letter of
\emph{American Journal of Physics}.

\subsection{Mean-field description and nonlinear dynamics of BEC}

From a theoretical standpoint, and for many experimentally relevant
conditions, the static and dynamical properties of a BEC can be described by
means of an effective mean-field equation known as the Gross-Pitaevskii (GP) equation \cite{PethickSmith,PitStrin}. This is a variant of
the famous nonlinear Schr{\"o}dinger equation (NLSE), incorporating an
external potential used to confine the condensate; NLSE is known to be a
universal model describing the evolution of complex field envelopes in
nonlinear dispersive media \cite{dodd,sulem,ablowitz1}. In the case of BECs,
the nonlinearity in the GP (NLSE) model is introduced by the interatomic
interactions, accounted for through an effective mean field. Thus, an
inherent feature of the BEC dynamics is its \textit{nonlinearity}, which is
induced by collisions between atoms, in spite of the fact that the density
of the quantum bosonic gases is very low.

The studies of the matter waves in the presence of the nonlinearity drive a
vast research area known as ``nonlinear atom optics'' (see, e.g., Refs.~\cite%
{nlatom}).
Importantly, many collective excitations, including self-trapped localized
states supported by the condensate's intrinsic nonlinearity (e.g.,
solitons), are less straightforward to create (and difficult to describe by
adequate models) in dense media featuring macroscopic quantum phenomena,
such as liquid helium. An exception are phase solitons (fluxons), i.e.,
quanta of magnetic flux trapped in long Josephson junctions formed by
superconductors \cite{7}, whose experimental and theoretical studies are
relatively straightforward and have been developed in detail \cite{52}.
Nevertheless, atomic BECs constitute an ideal setting for studies of such
macroscopic nonlinear excitations, as is explained in more detail below.

In the atomic BEC with intrinsic self-attraction (e.g., $^7$Li or $^{85}$Rb
BECs), the creation of effectively one-dimensional (1D) matter-wave \textit{%
bright solitons} (both isolated ones and multi-soliton sets) in \textit{%
cigar-shaped} configurations, which are tightly confined by external
potentials in the transverse plane, was successfully reported in condensates
of $^{7}$Li \cite{Randy,Paris,Randy2} and $^{85}$Rb \cite{Weiman,Billam}
(see also the reviews in Ref.  \cite{in-book}). Collisions between moving quasi-1D
solitons also admit accurate experimental implementation \cite{Jason} and
theoretical analysis \cite{Cornish,Jason}.

More typical is the repulsive sign of the inter-atomic interactions (as in
the cases of $^{87}$Rb or $^{23}$Na BECs), which lends the BEC the effective
self-repulsive nonlinearity. This kind of the intrinsic interaction readily
creates \textit{dark solitons}, which were predicted theoretically \cite%
{dark-prediction} and created experimentally \cite{dark,dark2} in BECs loaded
into a cigar-shaped trap. In fact, dark solitons were first created \cite%
{dark} prior to the realization of the above-mentioned bright solitons in
the self-attractive condensates, placed into the same type of the trapping
potential (a review of the topic of dark solitons in BEC was given in
article \cite{Dimitri}). Similar to the case of bright solitons, not only
single-soliton~\cite{dark} states, but also multiple dark solitons were
created \cite{dark2}, while their interactions and collisions were also
studied both in theory \cite{dark2,Huang} and experiments \cite%
{dark2,hambcol}.

Later, stable \textit{dark-bright soliton} complexes in binary BEC were
predicted in theory \cite{buschanglin} and observed in experiments \cite%
{dark-bright} as well. In more recent works, multiple dark-bright solitons
\cite{pe1}, as well as \textit{dark-dark solitons} \cite{pe2} were also
experimentally created. In addition, in the same setting of the
self-repulsive nonlinearity, not only dark solitons, but also bright
solitons are possible too: in particular, if -- instead of the usual
parabolic trap -- a periodic (optical lattice) potential \cite{Morsch} is
used to confine the condensate, then \textit{gap solitons} can be formed.
Experimentally, matter-wave gap solitons built of $\sim 250$ atoms in a $%
^{87}$Rb condensate, were reported in Ref.~\cite{Markus}.

On the other hand, in the case of multidimensional BEC geometry, there has
been an intense theoretical and experimental activity on vortices and vortex
structures in BECs with the self-repulsive intrinsic nonlinearity (see Refs.~%
\cite{9} for reviews on this topic). This is due to the fact that vortices
are intimately related to the superfluid properties of BECs, and play an
important role in transport, dissipative dynamics and quantum turbulence
(see, e.g., Refs.~\cite{turb}). Historically, the first observation of
vortices in BECs was achieved by \textit{phase imprinting} \cite{8} (a
technique that was also used to create dark solitons \cite{dark}). Nevertheless,
there exist other techniques that have been used in experiments to nucleate
vortices in BECs: these include \textit{stirring} the condensate above a
certain critical angular speed \cite{Madison00} (this method was used to
create \textit{vortex lattices} \cite{vlat}), \textit{nonlinear interference}
between different condensate fragments \cite{bpastuff} (this technique was
also employed for the creation of dark solitons \cite{dark2}), by forcing
superfluid flow around a repulsive Gaussian obstacle within the BEC \cite%
{bpavd}, and through the \textit{Kibble-Zurek mechanism} \cite{bpanat} (the
latter was originally proposed for the formation of large-scale structures in
the universe by means of a quench through a phase transition \cite{KZ}).

%

It is also important to mention that there still exist other notable results
concerning nonlinear phenomena in BECs. These include the realization of
few-vortex clusters and complex structures such as vortex dipoles, vortex
tripoles, parallel vortex rings, etc., and the study of their dynamics \cite%
{Madison00,bpavd,dshstuff1,dshstuff2,dshstuff3,vd1,vd2,vd3,vd4,vd5,vd6,vd7,vd8,vd9,vd10,vd11}%
, the observation of quantum shock waves \cite{hau}, the realization and
study of ``hybrid'' soliton-vortex structures \cite{sovo}, the observation
of Josephson oscillations~\cite{65} and spontaneous symmetry breaking
transitions~\cite{zibold} in BECs loaded into a symmetric double-well
trapping potential, and so on.

Many of the nonlinear phenomena mentioned above can be successfully studied
in the framework of mean-field theory. Nevertheless, there exist other
phenomena (where the underlying intrinsic BEC nonlinearity is also
important) that can not be described by means of the mean-field theory.
Examples of such situations, along with cases where quantum and/or thermal
fluctuations come into play, will be discussed below.

\subsection{Why are Bose-Einstein condensates always ``in vogue''?}


A great advantage offered by the BEC in low-density atomic gases is that
these media can be easily and very efficiently controlled by means of
external magnetic and optical fields. This circumstance enables various
experiments, and provides a framework for very accurate theoretical models.
As a result, the ultracold gases can be used for \textit{emulation} of many
phenomena which originate, e.g., in condensed matter physics, but take a
very complex form in the original settings, due to the strong interactions
in them, while in atomic BECs similar effects can be simulated in a much
simpler (``clean'') form.

An example which has recently drawn a great deal of interest is the
emulation of the \textit{spin-orbit coupling} (SOC), i.e., the interaction
between the motion of an electron or hole and their spin, or, in other
words, the linear mixing between two components of the electron's or hole's
spinor wave function. It is a fundamentally important phenomenon in
semiconductors, known in two distinct forms, as the \textit{Dresselhaus} and
\textit{Rashba} SOC \cite{semi}. In binary atomic BECs, created as a mixture
of two different hyperfine states of atoms of $^{87}$Rb, the SOC has been
experimentally implemented as a similar linear coupling between the two
atomic components, with the help of appropriate laser beams illuminating the
condensate, and a dc magnetic field applied to it \cite{Spielman}.

There are many other prominent examples of such an emulation provided by the
atomic-gas BEC. One such example concerns the direct observation of the
transition between the superfluid and Mott-insulator states formed in the
BEC loaded into an ``egg-carton'' periodic potential for individual atoms,
which is, in turn, imposed by an optical-lattice structure \cite{Greiner}. Other examples include resonantly-enhanced
tunneling in periodic potentials \cite{RET}, artificial gauge fields \cite%
{Hall,gauge}, topological insulators \cite{gauge}, the fractional quantum
Hall effect \cite{Hall}, and so on. Recently, the general topic of using the
BEC in atomic gases as the ``quantum simulator'', and the related theme of
using ultra-cold fermionic gases for the same general purpose, have been
reviewed in a series of articles \cite{simulators} and in a recent  book \cite{Lewenstein}.


As concerns applications of BEC, the most straightforward one is, arguably,
the use of matter waves for interferometry. With the help of the intrinsic
nonlinearity, they may feature superb accuracy, in comparison with
traditional optical interferometric schemes \cite{72}. Especially promising
are interferometric schemes of the Mach-Zehnder type, based on splitting and
subsequent recombination of bright matter-wave solitons \cite%
{Cuevas,MZ,in-book}. More generally, nonlinear waves that arise in BECs play a principal role in different
applications,  in
addition to their intrinsic interest. For instance, bright solitons have been argued to provide the
potential for 100-fold improved sensitivity for interferometers and their
lifetime of a few seconds enables precise force sensing applications~\cite%
{markk} (see also Ref.~\cite{martin}). For repulsively interacting BECs too,
the inter-atomic interactions have also been suggested to increase the
sensitivity to phase shifts, precisely due to the emergence of dark solitons
which enable (e.g.~through their oscillation in the confining trap) better
detection schemes~\cite{negretti}. Furthermore, nonlinear atom
interferometers can overcome the limitations of the current
state-of-the-art standard 
based on the 
so-called Ramsey spectroscopy~\cite{markus1,markus2}, due to their ability
to surpass the classical precision limit. Finally, vortices present their
own potential for applications. An example is the so-called ``analogue
gravity''~\cite{analogue}, whereby they may play a role similar to spinning
black holes. This allows to observe, in 
experimentally controllable environments, associated phenomena such as the
celebrated Hawking radiation or even simpler ones such as the super-radiant
amplification of sonic waves scattered from black holes~\cite{savage}.

BECs have also been experimentally demonstrated to be usable for performing
remarkable tasks such as the implementation of the Gaussian sum algorithm
for factoring numbers~\cite{sadgrove}, by exploiting higher order quantum
momentum states, improving in this way the algorithm's accuracy, once again,
beyond its classical implementation. This is in line with the development of
the Shor algorithm as an efficient quantum mechanical way to factorize large
numbers, a task thought to be classically intractable~\cite{sadgr3,sadgr4}.

Another potentially promising application is the use of BEC as a resource
for the implementation of quantum computing. In this context, one
possibility is to use trapped droplets of the condensate as qubits \cite%
{qubit,JakschZoller}. In optical lattices also, atomic analogs of
semiconductor electronic circuits (the so-called ``atomtronics'') have been
proposed, in order to realize quantum devices such as diodes and transistors~%
\cite{holland}. On the other hand, collision between quantum matter-wave
solitons can be used to induce entanglement between them \cite{Maciek}, which
is a prerequisite necessary for the design of soliton-based
quantum-information-processing schemes.

Also promising is the development of atom-wave lasers, which should be able
to emit high-intensity coherent matter-wave beams, in continuous-wave (CW)
or pulsed (soliton-like) regimes. Such beams may be very useful, in
particular, for precision measurements. The first experimental realization
of a CW matter-wave laser was reported in Ref. \cite{laser1}, which was
followed by the development of a design with a separated BEC\ reservoir and
the beam-emitting cavity \cite{laser2}. These experimental works used 
condensates consisting of $^{87}$Rb atoms. A review of experimental results on the topic
of matter-wave lasers was recently published in Ref. \cite{laser-review}. Models
of matter-wave lasers operating in a pulsed regime were developed
theoretically \cite{laser3}.

\section{Models, settings and basic results}



\subsection{The Gross-Pitaevskii equation}

As mentioned above, the fundamental model which provides for an accurate
description of the BEC in dilute degenerate gases of bosonic atoms is based
on lowest-order mean-field theory. According to this approach, the gas is
described by means of the Gross-Pitaevskii equation (GPE) for the
single-particle wave function, $\Psi \left( x,y,z,t\right) $, where $\left(
x,y,z\right) $ and $t$ are the coordinates and time [as mentioned above,
``degenerate'' means that the de Broglie wavelength of atoms moving with the
thermal velocity in the dilute gas is large enough in comparison with the
mean inter-atomic distance -- see Eq.~(\ref{degen})]. The three-dimensional
(3D) form of the GPE is written, in physical units, as:
\begin{equation}
i\hbar \frac{\partial \Psi }{\partial t}=-\frac{\hbar ^{2}}{2m}\left( \frac{%
\partial ^{2}}{\partial x^{2}}+\frac{\partial ^{2}}{\partial x^{2}}+\frac{%
\partial ^{2}}{\partial x^{2}}\right) \Psi +U\left( x,y,z;t\right) \Psi +%
\frac{4\pi \hbar ^{2}}{m}a_{s}\left( x,y,z;t\right) \left\vert \Psi
\right\vert ^{2}\Psi ,  \label{GPE}
\end{equation}
where $m$ is the atomic mass, $U\left( x,y,z;t\right) $ is the external
potential acting on individual atoms and thus confining the condensate as a
whole ($U$ may depend on time too, which is often called \textit{management}
of the potential \cite{management}), and $a_{s}\left( x,y,z;t\right) $ is
the scattering length which determines collisions between the atoms: $%
a_{s}>0 $ and $a_{s}<0$ correspond to repulsive and attractive interactions,
respectively. The spatial and temporal dependence of $a_{s}$, which is
important for many predictions of nontrivial dynamical states in the BEC
(see below), may be induced by means of the\textit{\ Feshbach-resonance}
(FR) \textit{management }technique \cite{management}. FR implies the
formation of a quasi-bound state of two atoms in the course of the collision
between them in the presence of an external magnetic field \cite{Inouye}, or
under appropriate laser illumination \cite{Killian}. The FR can be induced
too by combined magneto-optical settings \cite{magneto-optical}. Making use
of the very accurate tunability of the FR \cite{tunable}, spatially non-uniform
and/or temporally variable external fields, controlling the FR, can be
employed to induce spatially and/or time-dependent nonlinearity
coefficients, accounted for by $a_{s}\left(x,y,z;t\right) $. Further, the
wave function is subject to the normalization condition, which is determined
by the total number, $N$, of atoms in the condensate:
\begin{equation}
\int \int \int \left\vert \Psi \left( x,y,z;t\right) \right\vert^2 dxdydz=N.
\label{N}
\end{equation}
Note that, alternatively, the wave function may be defined with the unitary
norm, $\int \int \int \left\vert \Psi \left( x,y,z;t\right) \right\vert^2
dxdydz=1$, replacing $a_{s}$ in Eq. (\ref{GPE}) by $Na_{s}$.

The GPE and its variants constitute a relatively simple mathematical
framework, which admits precise simulations and the use of effective
analytical approaches. The latter include the Thomas-Fermi approximation
(TFA), which neglects the terms of the second derivative (the kinetic energy of the
quantum particles) in Eq.~(\ref{GPE}) \cite{PitStrin}, and a more accurate
and versatile variational approximation, which has found a great number of
applications to BECs \cite{42,43}. The TFA is relevant for $a_{s}>0$ (the
self-repulsive nonlinearity), for solutions which, in the simplest case, do
not include a nontrivial phase structure; in such a case, the solution with
chemical potential $\mu >0$ is found by using the ansatz $\Psi =e^{-i\mu
t}\Phi_{\mathrm{TFA}}(x,y,z)$, where the density $\left| \Phi _{\mathrm{TFA}%
}(x,y,z)\right|^{2}$ is approximated as:
\begin{equation}
\left| \Phi _{\mathrm{TFA}}(x,y,z)\right|^{2}=\frac{m}{4\pi \hbar
^{2}a_{s}\left( x,y,z\right) }\left\{
\begin{array}{c}
\mu -U\left( x,y,z\right) ,~~\mathrm{for}~U\left( x,y,z\right) <\mu ; \\
0,~~\mathrm{for}~U\left( x,y,z\right) >\mu .%
\end{array}%
\right.  \label{Thomas}
\end{equation}
The TFA may be easily generalized for vortex states, which are sought for as
solutions to Eq. (\ref{GPE}) in the cylindrical coordinates $( \rho \equiv
\sqrt{x^{2}+y^{2}},~\theta ,~z) $, assuming that $U=U\left( \rho ,z\right) $
and $a_{s}=a_{s}\left( \rho ,z\right) $ are subject to the cylindrical
symmetry:
\begin{equation}
\Psi =e^{-i\mu t+iS\theta }\Phi \left( \rho ,z\right) ,  \label{vortex}
\end{equation}
with real \emph{integer} vorticity $S$, and 
function $\Phi $ satisfying the respective stationary equation:
\begin{equation}
\mu \Phi =-\frac{\hbar ^{2}}{2m}\left( \frac{\partial ^{2}}{\partial \rho
^{2}}+\frac{1}{\rho }\frac{\partial }{\partial \rho }-\frac{S^{2}}{\rho ^{2}}%
+\frac{\partial ^{2}}{\partial z^{2}}\right) \Phi +U\left( \rho ,z\right)
\Phi +\frac{4\pi \hbar ^{2}}{m}a_{s}\left( \rho ,z\right) |\Phi|^{2}\Phi.
\label{Phi}
\end{equation}
The TFA neglects all the terms with $\rho $- and $z$-derivatives in Eq. (%
\ref{Phi}), which yields the following generalization of solution (\ref%
{Thomas}) \cite{Feder,gyroscope}:
\begin{equation}
\left| \Phi _{\mathrm{TFA}} \right|^{2}=\frac{m}{4\pi \hbar ^{2}a_{s}\left(
\rho ,z\right) }\left\{
\begin{array}{c}
\mu -\left[ U\left( \rho ,z\right) +\frac{\hbar ^{2}S^{2}}{2m}\rho ^{-2}%
\right] ,~~\mathrm{for}~U\left( \rho ,z\right) +\frac{\hbar ^{2}S^{2}}{2m}%
\rho ^{-2}<\mu ; \\
0,~~\mathrm{for}~U\left( \rho ,z\right) +\frac{\hbar ^{2}S^{2}}{2m}\rho
^{-2}>\mu .%
\end{array}%
\right.
\end{equation}

Techniques for numerical treatment of GPEs have been developed in great
detail too. They include methods for the solution of boundary-value
problems, aimed at finding stationary states trapped in external potentials,
or self-trapped due to the nonlinearity (solitons), as well as direct
simulations of the GPE in real or imaginary time (the latter approach helps
to generate stationary solutions for ground states \cite{46}, due to the
relaxational character of the dynamics). In particular, the
semi-implicit split-step Crank-Nicolson algorithm \cite{CN} has become a method
of choice for solving GPEs in many settings. The unconditional
stability of this algorithm makes it especially useful in studies of
real-time dynamics of BECs, although it can be equally well used to
produce ground states of relevant experimental setups, including
fast-rotating BECs with many vortices. The readily available Fortran and C
codes \cite{AB-GP}, which implement the Crank-Nicolson approach in 1D, 2D, and
3D geometries of BECs with different symmetries, are well tested and highly
optimized. Furthermore, the C codes are parallelized using the OpenMP
approach, which speeds up execution of numerical simulations of BECs
significantly, up to one order of magnitude on modern computers with
multi-core CPUs. Apart from the imaginary-time propagation implemented in the
framework of GPEs, ground states of a BEC (and of other quantum systems described
by linear and nonlinear Schr{\"o}dinger equations) can also be calculated
numerically using the higher-order effective-action approach \cite{AB-EA}, as
demonstrated in Refs.~\cite{AB-EA-APP}. Other popular methods used to solve
the GPE by means of pseudospectral and finite-difference methods are detailed in
Refs. \cite{GPEA0,GPEA1,GPEA2,GPEA3}. A survey of analytical and numerical
methods used in the studies of BEC models was given in review article
\cite{nonlinearity}.

For the BEC composed of chromium \cite{Pfau} or dysprosium \cite{Lev} atoms,
with magnetic moments, $\mu $, polarized in certain direction by an external
uniform dc magnetic field, the GPE includes a nonlinear term
accounting for the long-range dipole-dipole interactions:
\begin{equation}
i\hbar \frac{\partial \Psi }{\partial t}=-\frac{\hbar ^{2}}{2m}\nabla
^{2}\Psi +U\left( \mathbf{r};t\right) \Psi +\frac{\mu _{0}\mu ^{2}}{4\pi }%
\Psi (\mathbf{r})\int \left\vert \Psi (\mathbf{r}^{\prime })\right\vert ^{2}%
\frac{1-3\cos ^{2}\theta }{\left\vert \mathbf{r}-\mathbf{r}^{\prime
}\right\vert ^{3}}d\mathbf{r}^{\prime }.  \label{DD}
\end{equation}
In Eq.~(\ref{DD}), $\nabla ^{2}$ stands for the 3D Laplacian, the term $\sim
a_{s}$, which represents the usual contact interaction (see Eq. (%
\ref{GPE})), is dropped on grounds that it is small in comparison to the dipole-dipole interaction, $\mu _{0}$ is the magnetic permeability of vacuum, and $\theta $
is the angle between vector $\mathbf{r}-\mathbf{r}^{\prime }$ and the
polarization direction of the atomic magnetic moments.

Beyond the mean-field approximation, quantum fluctuations and interaction of
the condensate with the thermal component of the gas are described within
the framework of the Hartree-Fock-Bogoliubov equations, which are
essentially more cumbersome than the relatively simple GPE \cite{40}.
However, it is only under somewhat special conditions (i.e., small atom
numbers below $N \approx 1000$ or large enough temperatures, of the order of
many tens or hundreds nK) that fluctuations play a crucial role for coherent
matter-wave patterns, including solitons in the case of experiments.
Nevertheless, such settings are becoming of increasing interest in both
theoretical and experimental studies~\cite{ldc,prouk}.

\subsection{Two-component systems}

Shortly after the experimental realization of the single-component BEC,
advances in trapping techniques opened the possibility to simultaneously
confine atomic clouds in different hyperfine spin states. The first such
experiment, the so-called \textit{pseudospinor} condensate, was achieved in
mixtures of two magnetically trapped hyperfine states of $^{87}$Rb \cite%
{Myatt1997a}. Subsequently, experiments in optically trapped $^{23}$Na \cite%
{Stenger1998a} were able to produce multicomponent condensates for different
Zeeman sub-levels of the same hyperfine level, the so-called \textit{spinor
condensates}. In addition to these two classes of experiments, mixtures of
two different species of condensates have also been created by sympathetic
cooling (i.e. condensing one species and allowing the other one to condense
by taking advantage of the coupling with the first species) -- see examples
of such BEC mixtures below.

Regarding the modelling of multi-component BECs, it is natural to proceed
from the single-component GPE to the corresponding system of coupled GPEs.
In particular, for the simplest case of two-component mixtures, Eq.~(\ref%
{GPE}) is replaced by the following system of equations for mean-field wave
functions, $\Psi _{1}$ and $\Psi _{2}$, of the two components:
\begin{equation}
i\hbar \frac{\partial \Psi _{1}}{\partial t} =-\frac{\hbar ^{2}}{2m_{1}}
\nabla^{2} \Psi_{1} +U_{1}\left( \mathbf{r};t\right) \Psi _{1}+\frac{4\pi
\hbar ^{2}}{m_{1}}\left( a_{s}^{(1)}\left\vert \Psi _{1}\right\vert
^{2}+a_{s}^{(12)}\left\vert \Psi _{2}\right\vert ^{2}\right) \Psi_1,
\label{GPE1}
\end{equation}
\begin{equation}
i\hbar \frac{\partial \Psi _{2}}{\partial t} =-\frac{\hbar ^{2}}{2m_{2}}
\nabla^{2} \Psi _{2}+U_{2}\left( \mathbf{r};t\right) \Psi _{2}+\frac{4\pi
\hbar ^{2}}{m_{2}}\left( a_{s}^{(2)}\left\vert \Psi_{2}\right\vert ^{2}
+a_{s}^{(12)}\left\vert \Psi _{1}\right\vert ^{2}\right) \Psi_2,
\label{GPE2}
\end{equation}where $a_{s}^{(12)}$ is the scattering length for collisions between atoms
belonging to the two different species, for which the trapping potentials, $%
U_{1}$ and $U_{2}$, induced by the same external field, may be, generally
speaking, different. Masses $m_{1}$ and $m_{2}$ are different for BEC
mixtures composed by different atom species -- also-called \textit{%
heteronuclear} systems -- such as 
$^{85}$Rb$-$$^{87}$Rb \cite{85-87} and $^{87}$Rb$-$$^{133}$Cs \cite{hetero}
binary BEC mixtures. On the other hand, $m_{1}=m_{2}$ for mixtures of
different hyperfine states of the same atomic species; such mixtures were
experimentally realized for the first time with $^{87}$Rb atoms \cite{homo}.
In the case of repulsive intra- and inter-species interactions, when all
the scattering lengths in Eqs.~(\ref{GPE1}) and (\ref{GPE2}) are positive, that is, $%
a_{s}^{(1)}>0$, $a_{s}^{(2)}>0$, and $a_{s}^{(12)}>0$, the condition for the \textit{immiscibility} of the two
components -- and the onset of the separation between them -- is given by
\cite{Mineev}:
\begin{equation}
a_{s}^{(1)}a_{s}^{(2)}<\left( a_{s}^{(12)}\right) ^{2},  \label{immisc}
\end{equation}
which implies that the repulsion between atoms belonging to the different
components is stronger than the repulsion between atoms in each component.
Condition~(\ref{immisc}) pertains to the free infinite space, while the
pressure of the trapping potential makes the binary BEC more miscible,
shifting the critical point, $a_{s}^{(1)}a_{s}^{(2)}=\left(
a_{s}^{(12)}\right)^{2}$, to larger values of positive $a_{s}^{(12)}$ \cite%
{immisc-confinement}. Immiscible two-component condensates, loaded into a
trap, form domain walls separating the two components \cite{Poland}. Such
domain walls were observed in experiments \cite{DW-experiment}.

Numerical analyses of realistic trapped states of binary BECs have
revealed, in the case of cigar-shaped geometries, two distinct immiscible
stationary configurations: a segregated one, in which the two components face
one another, being separated by a domain wall,
and a symbiotic state, in which one component effectively traps
the other \cite{alexfar}. Both configurations do not directly obey the aforementioned
simple miscibility criteria, although they can be transformed into a
miscible configuration when the condensate is subject to a resonant drive
\cite{alexfar}. Finally, we mention that, while the symbiotic state is not a
soliton, numerous types of solitons are known to exist in spinor
condensates (see, e.g., \cite{SBin1,SBin2}).

In the case when the two wave-function components represent different
hyperfine states of the same atomic species, an external resonant
radiofrequency field (with frequencies in the GHz range) may add linear
mixing, with strength $\kappa $ (alias \textit{Rabi coupling}), to the
system, which is accounted for by extra terms $\kappa \Psi _{2}$ and $\kappa
\Psi _{1}$, added to equations (\ref{GPE1}) and (\ref{GPE2}), respectively
\cite{lincoup}. The interplay of the Rabi coupling with the repulsive
interactions causes a shift of the miscibility-immiscibility transition (\ref%
{immisc}) \cite{Merhasin} (see relevant experimental results in Ref.~\cite%
{mkobampgk}).

\subsection{Trapping potentials}

Coming back to the single GPE (\ref{GPE}), it is relevant to stress that two
most common types of the confining potential are the harmonic-oscillator
(HO)
\begin{equation}
U_{\mathrm{HO}} =\frac{m}{2}\left( \Omega _{x}^{2}x^{2}+\Omega
_{y}^{2}y^{2}+\Omega _{z}^{2}z^{2}\right)   \label{HO}
\end{equation}%
and optical-lattice (OL)
\begin{equation}
U_{\mathrm{OL}} =U_{x}\sin ^{2}\left( k_{x}x\right) +U_{y}\sin ^{2}\left(
k_{y}y\right) +U_{z}\sin ^{2}\left( k_{z}z\right)   \label{OL}
\end{equation}
ones, where $\Omega _{x,y,z}^{2}$ are \textit{trapping frequencies} of the
(generally, anisotropic) HO potential, and $U_{x,y,z}$ represent the depths
of the periodic OL potential. The OL is built as the classical interference
pattern by pairs of counterpropagating mutually coherent laser beams
illuminating the condensate, with respective wavelengths $\lambda
_{x,y,z}=2\pi /k_{x,y,z}$.

The OL is made attractive or repelling for individual atoms by red- or
blue-detuning of the illuminating light with respect to the frequency of the
dipole transition in the atoms. Typically, the OL wavelength $\lambda \sim 1$
$\mathrm{\mu }$m is used in experiments. Usually, the depth $U$ of the OL
potential is measured in natural units of the recoil energy, $E_{R}=\left(
\hbar k\right) ^{2}/\left( 2m\right) $. The use of the OL potentials for the
creation of matter-wave patterns in BEC was proposed in \cite{Jaksch} and
relevant applications, such as macroscopic quantum interference, immediately
ensued~\cite{andkas} -- see also reviews \cite{Bloch} and \cite{engineering}%
. A well-known example of the effect induced by the OL is the transition
from the bosonic superfluid to the Mott insulator \cite{Greiner}.

Generally, the technique based on the use of OLs is similar to that which
was proposed \cite{16} and implemented in the form of photonic lattices in
photorefractive media, producing a number of spectacular results, including
1D and 2D optical solitons and vortices in 2D \cite{vortex} -- see reviews
\cite{17}.

\subsection{The discrete system}

In both the BEC and photonic settings, a very deep (compared to the chemical
potential) OL potential effectively splits the mean-field matter-wave
function, or the optical electromagnetic field, into a set of nodes (each
one representing one well) weakly interacting between them via tunneling
coupling. 
In this case, using the expansion of the continuum field over a set of
Wannier modes localized around local wells, GPE (\ref{GPE}) can be reduced
to a discrete nonlinear Schr\"{o}dinger equation (NLSE) \cite%
{discrete,Panos-book}. In a properly scaled form, its 3D version is
\begin{eqnarray}
i\frac{d\Psi _{j,k,l}}{dt}&=&-\frac{1}{2}\left(\Psi _{j+1,k,l}+\Psi
_{j-1,k,l}+\Psi _{j,k+1,l}+\Psi _{k,k-1,l} \right.  \notag \\
& &+\left.\Psi_{j,k,l+1}+\Psi _{j,k,l-1}-6\Psi _{j,k,l}\right) - \left\vert
\Psi _{j,k,l}\right\vert ^{2}\Psi _{j,k,l},  \label{discre}
\end{eqnarray}
where $j,k,l$ are discrete coordinates on the lattice, and $\Psi _{j,k,l}$
are amplitudes of trapped matter-wave fragments, the 2D and 1D versions
being produced by obvious reductions of Eq. (\ref{discre}). The present form
of the discrete NLSE implies that the on-site nonlinearity is
self-attractive. However, unlike the continuous model, in the discrete one
the self-repulsive nonlinearity may be transformed into its self-attractive
counterpart by means of the well-known \textit{staggering transformation}
\cite{Panos-book}, $\Psi _{j,k,l}(t)\equiv \left( -1\right)
^{j+k+l}e^{-12it} \tilde{\Psi}_{j,k,l}^*(t)$, where the
asterisk stands for the complex conjugate.

The discrete NLSE gives rise to many species of solitons \cite{Panos-book},
especially interesting ones being \emph{discrete localized vortices}, which were
predicted theoretically \cite{MK} and created experimentally (as nearly
discrete objects) in photonics using waveguide arrays built in a
photorefractive material \cite{vortex}. Thus, while 1D and 2D versions of
Eq. (\ref{discre}) apply to the photonic settings \cite{17}, the 3D discrete
system is meaningful solely in the BEC context. In the latter case, it
generates complex stable localized modes, such as, e.g., \textit{discrete
Skyrmions} \cite{Skyrme}, diamonds, octupoles, oblique vortices, and vortex
cubes~\cite{pgkprl}, among many others.

The discrete NLSE suggests a direct transition to the fully quantum system
in the form of the Bose-Hubbard (BH) model, replacing the mean-field
(classical) lattice wave functions in Eq.~(\ref{discre}) by quantum
operators, $b_{j}$ (for simplicity, discussed here in the 1D setting). The
corresponding Hamiltonian is
\begin{equation}
H=\sum_{j}\left[ -J\,{b}_{j}^{\dagger }\left( {b}_{j+1}+{b}_{j-1}\right) +{%
\frac{1}{2}}U{n}_{j}({n}_{j}-1)\right] \;,  \label{ham}
\end{equation}
where ${n}_{j}={b}_{j}^{\dagger }{b}_{j}$ is the operator of the on-site
number of particles, $J$ is the inter-site-hopping constant, and $U$ is the
constant of the on-site interaction ($U>0$ and $U<0$ correspond, as before,
to the self-attraction and self-repulsion, respectively). A famous result
produced by Hamiltonian (\ref{ham})\ is the phase diagram separating the
quantum superfluid and Mott insulator (see Ref. \cite{Fisher}). In terms of
applications, the BH model is a natural tool for the theoretical analysis of
operations of BEC-based qubits. Reviews of the topic of BH in connection
to its realization in BEC can be found in articles \cite%
{BlochDalibardZwerger} and \cite{JakschZoller}. The well-elaborated
numerical technique for the analysis of the BH model and its modifications
is based on the density-matrix-renormalization-group method \cite{DMRG}.

\subsection{Reduction to lower-dimensional settings}

The above-mentioned nearly-1D cigar-shaped traps are represented by the
potentials which include tight confinement in the transverse plane, i.e.,
large $\Omega _{y}^{2}=\Omega_{z}^{2}\equiv \Omega _{\perp }^{2}$ in Eq.~(%
\ref{HO}), and an arbitrary weak potential, $U(x,t)$, acting in the axial
direction, $x$. In this case, the 3D wave function can be approximated by
the factorized \textit{Ansatz} (see, e.g., Ref.~\cite{vpg}),
\begin{equation}
\Psi \left( x,y,z,t\right) =\frac{1}{\sqrt{\pi }a_{\perp }}\exp \left(
-i\Omega _{\perp }t-\frac{y^{2}+z^{2}}{2a_{\perp }^{2}}\right) \psi \left(
x,t\right) ,  \label{factor}
\end{equation}
where the transverse part represents the ground state of the two-dimensional
HO in the transverse plane, with the respective oscillator length
\begin{equation}
a_{\perp }=\sqrt{\hbar /\left( m\Omega _{\perp }\right) }  \label{perp}
\end{equation}
(typical values relevant to the experiments are $a_{\perp }\simeq 3$ $%
\mathrm{\mu }$m), while the axial wave function, $\psi \left( x,t\right) $,
subject to the normalization condition $\int_{-\infty }^{+\infty }\left\vert
\psi \left( x\right) \right\vert ^{2}dx=N$, see Eq.~(\ref{N}), satisfies the
1D equation obtained by averaging in the transverse plane:
\begin{equation}
i\hbar \frac{\partial \psi }{\partial t}=-\frac{\hbar ^{2}}{2m}\frac{%
\partial ^{2}\psi }{\partial x^{2}}+U\left( x;t\right) \psi +\frac{4\hbar
^{2}}{ma_{\perp }^{2}}a_{s}\left( x;t\right) \left\vert \psi \right\vert
^{2}\psi,  \label{1D}
\end{equation}
which has the form of the one-dimensional NLSE, with an external potential, $%
U$. Essentially the same equation occurs in many other physical settings,
such as the light propagation in planar waveguides, in which case $t$ is
actually the propagation distance, while $-U(x)$ represents the confining
profile of the local refractive index \cite{KA,sulem,ablowitz1}.
Accordingly, Eq.~(\ref{1D}) with $a_{s}<0$ and $a_{s}>0$ gives rise,
respectively, to the commonly known bright- and dark-soliton solutions.
Interestingly, the NLSE in 1D is integrable in the case of $U=0$ and $a_{s}=%
\mathrm{const}$ \cite{Zakha}.

An interesting ramification of this setting is the toroidal quasi-1D trap,
which is described by Eq.~(\ref{1D}) with periodic boundary conditions in $x$%
. Such toroidal traps, realized by means of several different techniques,
are available in the experiment \cite{torus}.

A quasi-2D pancake-shaped (oblate) configuration, with strong confinement
acting in the transverse 1D direction, corresponds to large $\Omega
_{z}^{2}\equiv \Omega _{\perp }^{2}$ in Eq.~(\ref{HO}), combined with a
general relatively weak potential, $U\left( x,y\right) $, acting in the
pancake's plane. This configuration is approximated by the respective
factorized ansatz,
\begin{equation}
\Psi \left( x,y,z,t\right) =\frac{1}{\pi ^{1/4}\sqrt{a_{\perp }}}\exp \left(
-\frac{i}{2}\Omega _{\perp }t-\frac{z^{2}}{2a_{\perp }^{2}}\right) \psi
\left( x,y,t\right) ,
\end{equation}
which leads to the effective 2D equation,
\begin{equation}
i\hbar \frac{\partial \psi }{\partial t}=-\frac{\hbar ^{2}}{2m}\left( \frac{%
\partial ^{2}}{\partial x^{2}}+\frac{\partial ^{2}}{\partial y^{2}}\right)
\psi +U\left( x,y;t\right) \psi +\frac{2\sqrt{2\pi }\hbar ^{2}}{ma_{\perp }}%
a_{s}\left( x;t\right) \left\vert \psi \right\vert ^{2}\psi .
\label{pancake}
\end{equation}

The reduction of the 3D GPE (\ref{GPE}) to its 1D version (\ref{1D}) on the
basis of factorized \emph{Ansatz} (\ref{factor}) with the fixed transverse
localization radius, $a_{\perp }$, is relevant in the limit of low density.
For higher density, the reduction is also based on ansatz (\ref{1D}), in
which $a_{\perp }$ is allowed to be a variable parameter. Then, the
reduction to 1D is performed by means of the variational approximation,
which leads to a system of 1D equations for $\psi \left( x,t\right) $ and $%
a_{\perp }\left( x\right) $ \cite{Luca}, that can be reduced to a single
effective equation for the 1D wave function with a nonpolynomial
nonlinearity. 
The resulting ``nonpolynomial NLSE'' (abbreviated as NPSE \cite{Luca}), and
the respective local expression for $a_{\perp }$, are given (in a scaled
form) by

\begin{equation}
i\frac{\partial \psi }{\partial t} =-\frac{1}{2}\frac{\partial ^{2}\psi }{%
\partial x^{2}}+U\left( x,t\right) \psi +\frac{1+\left( 3/2\right) g|\psi
|^{2}}{\sqrt{1+g|\psi |^2}}\psi ,  \label{NPSE}
\end{equation}
and
\begin{equation}
a_{\perp }^{4} =1+g|\psi |^{4},~g\equiv 2a_{s}/a_{\perp }<0.  \label{g}
\end{equation}
%
However, the relevant reduction from 3D to 1D may be done in multiple ways
(e.g., working at the level of underlying Lagrangian/Hamiltonian structure
and of the corresponding action or at that of the equations of motion). On a related direction, using the standard adiabatic approximation and accurate results for the local chemical potential,  one obtains an alternative equation with a nonpolynomial
nonlinearity~\cite{Canary}:
\begin{equation}
i\frac{\partial \psi }{\partial t}=-\frac{1}{2}\frac{\partial ^{2}\psi }{%
\partial x^{2}}+U\left( x,t\right) \psi +\sqrt{1+2g\left\vert \psi
\right\vert ^{2}}\psi ,  \label{Del}
\end{equation}%
where $g$ is the same as in Eq. (\ref{g}).

\subsection{Collapse of attractive condensates}

In the free space ($U=0$), with a constant negative scattering length,
corresponding to the self-attractive nonlinearity ($a_{s}<0$), a rescaled
form of Eq.~(\ref{pancake}) amounts to the 2D version of the NLSE:
\begin{equation}
i\frac{\partial \psi }{\partial t}=-\frac{1}{2}\left( \frac{\partial ^{2}}{%
\partial x^{2}}+\frac{\partial ^{2}}{\partial y^{2}}\right) \psi -\left\vert
\psi \right\vert ^{2}\psi.  \label{2D}
\end{equation}
A well-known fact is that Eq. (\ref{2D}) gives rise to a family of
isotropic, so-called \textit{Townes' solitons} \cite{Townes},
\begin{equation}
\psi =\exp \left( -i\mu t\right) \phi (r), \quad r\equiv \sqrt{x^{2}+y^{2}},
\label{phi}
\end{equation}
%
with arbitrary chemical potential $\mu <0$, and real function $\phi $
obeying the equation:
\begin{equation}
\mu \phi +(1/2)\left( \phi^{\prime \prime }+r^{-1}\phi ^{\prime }\right)
+\phi ^{3}=0.
\end{equation}
The family of the Townes' solitons is degenerate, in the sense that their
norm takes a single value which does not depend on $\mu $:
\begin{equation}
N_{\mathrm{T}}=2\pi \int_{0}^{\infty }\phi ^{2}(r)rdr\approx 5.85.
\label{NT}
\end{equation}
Note that an analytical variational approximation for this norm predicts $N_{%
\mathrm{T}}=2\pi $, with relative error $\approx 7\%$ \cite{Anderson}.

On the other hand, the three-dimensional GPE (\ref{GPE}) in the free space,
with the uniform self-attractive nonlinearity, $U=0$ and $a_{s}<0$, gives
rise to a family of isotropic soliton solutions in the form given by Eq.~(%
\ref{phi}). 
Unlike their 2D counterparts in the form of the above-mentioned Townes'
solitons, the norm of the 3D solitons depends on $\mu $, $N=\mathrm{const}%
\cdot \left( -\mu \right) ^{-1/2}$, cf. Eq.~(\ref{NT}). The celebrated
Vakhitov-Kolokolov (VK) necessary stability criterion \cite{VK,Berge}, $%
dN/d\mu <0$, does not hold for this $N(\mu )$ dependence, hence the entire
family of the 3D free-space solitons is \emph{unstable}, which is completely
corroborated by the full analysis of the stability \cite{Berge}. For the 2D
Townes solitons, Eq. (\ref{NT}) formally predicts neutral VK stability, $%
dN/d\mu =0$, but in reality the Townes solitons are unstable too.
However, their instability is nonlinear, i.e., it is not accounted form by
any unstable eigenvalue in the spectrum of eigenmodes computed around the
stationary soliton, using the respective Bogoliubov-de Gennes equations (BdGEs). The eigenvalue associated
with the instability in this special case is at $0$, corresponding to a
special invariance arising in this critical case, namely the so-called
conformal invariance~\cite{sulem}, which allows a rescaling of the solitary
wave. In fact, the instability of the multidimensional solitons is explained
by the fact that the NLSE  with the self-attractive cubic
nonlinearity gives rise to the dynamical \textit{collapse}, i.e., the
self-similar formation of a true singularity after a finite evolution time.
In the 2D space, the collapse is \textit{critical}, which implies, \textit{%
inter alia}, that the norm of collapsing solutions must exceed a threshold
(minimum) value, which is precisely the Townes-soliton norm (\ref{NT}),
while the 3D collapse is supercritical, as its threshold norm is zero \cite%
{Berge}. In the experiments with the self-attractive BEC, the onset of the
collapse was readily observed (the first time in $^{85}$Rb \cite%
{collapse-exper}), as spontaneous explosion of the condensate (the so-called
``Bose nova''). It is interesting to mention that the small part of the
condensate surviving the explosion, can form a stable soliton in $^{85}$Rb
\cite{Weiman}.

\subsection{Models for non-BEC ultracold gases}

The above discussion pertains to quantum bosonic gases. Ultracold
fer\-mi\-onic gases have also been created in the experiment \cite{Jin1}, which
was followed by the observation of their condensation into the bosonic gas
of Bardeen-Cooper-Schrieffer (BCS) pairs \cite{Jin2}. The theoretical
description of the Fermi gases is more complex, because, in the general
case, the Pauli principle prevents the application of the mean-field
approximation to fermions, making it necessary to treat such gases directly
as systems with many degrees of freedom of individual particles (see, e.g.,
the works~\cite{Salerno} and the review \cite{FERMI}).

An approach to sufficiently dense Fermi gases is possible in terms of a
hydrodynamic description, which, in a sense, is a variety of the mean-field
theory. This approach starts from the famous works by Yang and Lee who
derived the energy density for a weakly coupled BCS superfluid \cite{Yang}.
In the spirit of the hydrodynamic approach, an effective equation for an
order parameter of the Fermi gas, $\Psi \left( x,y,x,t\right) $ was derived,
which seems as the NLSE with the self-repulsive term of power $7/3$ \cite%
{Adhikari}:
\begin{equation}
i\hbar \frac{\partial \Psi }{\partial t}=-\frac{\hbar ^{2}}{2m_{\mathrm{eff}}%
} \nabla^2 \Psi +U\left( x,y,z\right) \Psi +\frac{\hbar ^{2}}{2m}\left\vert
\Psi \right\vert ^{4/3}\Psi ,  \label{4/3}
\end{equation}
where $m_{\mathrm{eff}}$ is the effective mass, which may be different from
the particle's mass. This equation is valid for a slowly varying
order-parameter field, under the condition that the local Fermi energy is
much larger than all other local energy scales, such as potential $U\left(
x,y,z\right) $. Equation (\ref{4/3}) and its 1D and 2D reductions can be
used to predict various stationary and quasi-stationary density patterns in
the Fermi gas \cite{Adhikari}. Related equations for the dynamics of Fermi gases are discussed in Ref. \cite{FermiSal}.

Similarly to the case of bosonic gases, nonlinear excitations of Fermi gases
have attracted attention. In particular, dark solitons in a Fermi gas were
predicted near the BEC-BCS transition, using the description in terms of the
BdGEs \cite{dark-Fermi}, while their dynamical properties were studied in
several works \cite{dark-F2}. It is worth noting that dark solitons \cite%
{dark-Fexp} and hybrid soliton-vortex structures~\cite{Fsolvo} (the
so-called solitonic vortices) were recently observed in experiments with
Fermi gases in the BEC-BCS crossover.

Another well-known example of a dilute quantum medium different from BEC is
the 1D Tonks-Girardeau (TG) gas, formed by hard-core bosons, a solution for
which may be mapped into that for a gas of non-interacting fermions \cite%
{Fermi-TG}. In particular, this mapping (also known as ``Bose-Fermi
mapping'') makes it possible to produce a solution for a dark soliton in the
TG gas \cite{dark-TG}. Importantly, the TG gas was realized experimentally
using ultracold $^{87}$Rb atoms loaded into a tightly confined quasi-1D trap
\cite{TG-experiment}.

An equation for the order parameter of the TG gas was derived in work \cite%
{Kolo}, in the form of the 1D NLSE with the self-repulsive quintic term:
\begin{equation}
i\hbar \frac{\partial \Psi }{\partial t}=-\frac{\hbar ^{2}}{2m}\frac{%
\partial ^{2}\Psi }{\partial x^{2}}+U\left( x\right) \Psi +\frac{\pi \hbar
^{2}}{2m}\left\vert \Psi \right\vert ^{4}\Psi .  \label{quint}
\end{equation}
Similar to Eq. (\ref{4/3}), Eq.~(\ref{quint}) is valid only as a
quasi-stationary one and it does not provide correct description of
dynamical effects in the TG gas \cite{critical}. There exist examples of a
relevant use of Eq.~(\ref{quint}), including the prediction of solitons
supported by the long-range dipole-dipole attraction between atoms forming
the gas \cite{BBB-DD}, and the study of dark soliton oscillations \cite{fpk}%
. Interestingly, the result for the soliton oscillation frequency, which was
analytically found in Ref.~\cite{fpk} to be equal to the axial trap
frequency, was in agreement with numerical predictions obtained in Ref.~\cite%
{buhuy} via the Bose-Fermi mapping.

\section{Some special topics that have attracted interest}

\subsection{Temporal and spatial management}

The \textit{management} concept can be applied for the trapping potentials
by making the HO or OL strengths time-dependent. A typical example of
results produced by the time-periodic management of the HO potential (\ref%
{HO}), with $\Omega ^{2}=\Omega _{0}^{2}+\Omega _{1}^{2}\sin \left( \omega
t\right) $ (in the simplest 1D setting), is the prediction of parametric
resonances of self-trapped matter-wave packets (solitons) in the latter
setting \cite{BBB}. On the other hand, for the OL the management was
experimentally realized \cite{shaking} in the form of a ``rocking" OL, by
introducing a small wavelength mismatch between the two laser beams building
the OL, which is made a periodic function of time: $\Delta \lambda =\Delta
\lambda _{0}\sin \left( \omega t\right) $, i.e.,
the lattice as a whole performs periodic oscillatory motion. In particular,
the rocking OL potential may effectively suppress the matter-wave tunneling
across the lattice \cite{shaking}.

One particular application of the periodic modulation of the
strength of the HO potential deals with the appearance of regular patterns in
the density profile of the condensate through a modulational instability.
A prototypical example is the emergence of Faraday patterns in cigar-shaped
BECs \cite{Nicolin1} through the periodic modulation of the strength of the
radial component of the magnetic trap, similar experimental results
being known for $^4$He \cite{Nicolin2}. As another aspect of theoretical results,
we mention numerous studies of Faraday waves in models of the condensates
with short-range interactions \cite{Nicolin3,Nicolin3b,Nicolin4}, dipolar
condensates \cite{Nicolin5}, binary condensates with short-range
interactions \cite{Nicolin6}, Fermi-Bose mixtures \cite{Nicolin7}, and
superfluid Fermi gases \cite{Nicolin8}. As a related topic, let us mention
that it has been shown theoretically that Faraday waves can be suppressed in
condensates subject either to resonant parametric modulations \cite{Nicolin9}
or space- and time-modulated potentials  \cite{Nicolin10,Nicolin11}, and that
pattern-forming modulational instabilities lead to chaotic density profiles
\cite{Nicolin3} akin to those of turbulent BECs \cite{turb, NicolinB2}.
Apart from the use of parametric excitations, the formation of density  waves
has been studied in expanding ultra-cold Bose gases (either fully \cite{Nicolin12}
or only partly condensed \cite{Nicolin13,Nicolin14}), and the spontaneous
formation of  density waves has been reported for antiferromagnetic BECs
\cite{Nicolin15}.

As predicted theoretically (see Ref. \cite{RMP} for a review), many possibilities
for the creation of matter-wave patterns in BEC are offered by various
patterns of spatial \cite{NL-theory} and temporal \cite{Ueda-Fatkh}
modulation of the local scattering length, $a_{s}$, as implied by the
general form of GPE\ (\ref{GPE}). Experimentally, spatial ``landscapes'' of
the scattering length can be induced, via the FR mechanism, by the
corresponding spatially periodic distributions of the magnetic field (%
\textit{magnetic lattices}), created with the help of periodic structures
built of ferromagnetic materials \cite{magn-latt}. Another possibility is
the local modulation of the scattering length, which may be imposed, via the
optical FR, by time-average patterns ``painted'' by rapidly moving laser
beams \cite{painting}. Spatial modulation of
interatomic interactions has also been demonstrated at the submicron level
via pulsed optical standing waves in an ytterbium BEC~\cite{takahashi}. Once
again in this context, it is relevant to point out that some of these
possibilities, such as the temporal modulation of the nonlinearity have been
also realized in parallel, in other contexts such as, e.g., nonlinear optics~%
\cite{psaltis}.

Employing such magnetically or optically induced Feshbach resonances, via
the above-mentioned FR-management technique, indeed constitutes one of the
most promising methods for manipulating BECs. Such a possibility to control
the effective nonlinearity of the condensate, has given rise to many
theoretical and experimental studies. Probably the most well-known example
between these, is the formation of bright matter-wave solitons and soliton
trains in attractive condensates \cite{Randy}-\cite{Billam}, by switching
the interatomic interactions from repulsive to attractive.

This inspired many theoretical works studying the BEC dynamics under
temporal and/or spatial modulation of the nonlinearity. In particular, the
application of FR-management technique, with the low-frequency modulation of
the strength of the magnetic field causing the nonlinearity to periodically
switch between attraction and repulsion, can be used to stabilize 2D
solitons in the free space \cite{Ueda-Fatkh} (in reality, a weak
two-dimensional HO trapping potential is necessary in the experiment).
However, this method does not work in 3D, nor for 2D vortex solitons. On the
other hand, the same technique can be used for the generation of robust
matter-wave breathers \cite{wefrm}.

On the other hand, the so-called ``collisionally inhomogeneous condensates''
(a term coined in Refs.~\cite{NL-theory}) controlled by the \textit{%
spatially modulated nonlinearity}, have been predicted to support a variety
of new phenomena. This new regime can be achieved by means of
magnetically or optically controlled Feshbach resonances.
The magnetic Feshbach resonances is a well-established experimental method,
which has been used to study the formation of ultracold molecules \cite{NicolinF1},
the BEC-BCS crossover \cite{NicolinF2}, and the production of Efimov trimer
states \cite{NicolinF3}, but the inhomogeneity length scale of the
necessary magnetic field is usually larger than the size of the atomic sample,
therefore this method is not very efficient in reaching the collisionally inhomogeneous
regime. However, the optical Feshbach resonance has been shown to allow fine
spatial control of the scattering length, and recent experimental results
demonstrate controllable modulations of the \emph{s}-wave scattering length
on the scale of hundreds of nanometers \cite{takahashi}.

The range of new nonlinear phenomena specific to the inhomogeneous-non\-li\-ne\-a\-ri\-ty
regime includes adiabatic compression of
matter waves \cite{NL-theory,salfkh}, Bloch oscillations of matter-wave
solitons \cite{NL-theory}, emission of the solitons and design of atom-beam lasers
\cite{vpglas}, dynamical trapping of matter-wave solitons \cite{borisgofx,giota}
enhancement of transmissivity of matter waves through barriers, \cite%
{giota,abgar}, creation of stable condensates exhibiting both attractive and repulsive
interatomic interactions \cite{borisgofx}, competition between
incommensurable linear and nonlinear lattices \cite{borgx}, the generation
of dark solitons and vortex rings \cite{darkgofx}, control of Faraday waves
\cite{alexfar2}, and many others. Importantly, the Feshbach resonance was used
in recent experiments to induce real spatial inhomogeneities of the
scattering length \cite{painting,takahashi}, which paves the way for implementation
of the above-mentioned phenomena in the experiment.

\subsection{Multidimensional localized structures}


\subsubsection{Attractive BEC}

The stabilization of multidimensional solitons is a problem of great
interest not only to BEC, but also to nonlinear optics and related research
areas \cite{review,review1,review2,review3,review4}. It was predicted
theoretically, but not as yet demonstrated experimentally, that the use of
OL potentials is a universal means for the stabilization of such solitons
\cite{20,20a} (a similar stabilization mechanism was predicted for 2D
optical solitons supported by the Kerr self-focusing nonlinearity in
photonic-crystal fibers \cite{66}). In particular, the stabilization of 2D
solitons is explained by the fact that the norm of such solitons, trapped in
the OL potential, takes values \emph{below} the threshold value (\ref{NT})
corresponding to the Townes' soliton, hence the solitons have no chance to
start collapsing. The same property, $N<N_{\mathrm{T}}$, explains the
stabilization of solitons trapped in the HO potential (\ref{HO}) \cite%
{HO-stabilization,HO-stabilization_2006}. For a comprehensive study of
stability of two-dimensional elliptic vortices in self-attractive
Bose-Einstein condensates, trapped by an anisotropic harmonic trapping
potential, see Ref. \cite{HO-stabilization_2010}.

On a different but related direction let us mention that OLs of the Bessel
type can support soliton rotation, see Refs. \cite%
{rotary1,rotary2,rotary3,rotary4} for the theoretical works and Ref. \cite%
{rotary5} for the experimental verification, a nonlinear phenomenon which
has been investigated in both BEC and nonlinear optical settings.

Still more challenging objects are multidimensional vortex solitons (i.e.,
self-trapped modes with embedded vorticity). In addition to the possibility
of the collapse, they are still more unstable to fragmentation by azimuthal
perturbations \cite{review,RMP}. An accurate analysis of the stability of
three-dimensional solitons with vorticity $S=1$ in self-attractive
Bose-Einstein condensates, trapped in an anisotropic three-dimensional HO
potential was reported in Ref. \cite{HO-stabilization_2007}. The analysis
predicts that vortex solitons can also be stabilized by OL potentials \cite%
{20}. Of course, the lattice breaks the axial symmetry, but, nevertheless,
the vorticity embedded into a localized state may be defined in this case
too \cite{20,21}. In the simplest form, stable vortex solitons with
topological charge (integer vorticity) $S$ can be constructed as $N$-peaked
ring-shaped patterns with the vorticity represented by the phase circulation
along the ring, with phase shift $\Delta\phi=2\pi/N$ between adjacent peaks
(see the straightforward definition (\ref{vortex}) of the vorticity for
isotropic settings). OLs may stabilize solitons even if the lattice's
dimension is smaller by one than the dimension of the physical space,
including 1D \cite{25} and 2D \cite{25,26} OLs in the 2D and 3D settings,
respectively.

A new approach to the creation of stable 2D solitons supported by the cubic
self-attraction, which was considered impossible until recently, was put
forward in theoretical work \cite{HS}. It is based on the system of two
GPEs, linearly coupled by first-derivative terms representing the
above-mentioned SOC of the Rashba type, and by nonlinear terms accounting
for collisions between atoms belonging to the two different atomic states,
which underlie the coupled system. In the scaled form, the system is given
by
\begin{equation}
i\frac{\partial \psi _{1}}{\partial t} =-\frac{1}{2}\nabla ^{2}\psi
_{1}-(|\psi _{1}|^{2}+\gamma |\psi _{2}|^{2})\psi _{1}+\lambda \left( \frac{%
\partial \psi _{2}}{\partial x}-i\frac{\partial \psi _{2}}{\partial y}\right)
\label{Ben1}
\end{equation}
and
\begin{equation}
i\frac{\partial \psi _{2}}{\partial t} =-\frac{1}{2}\nabla ^{2}\psi
_{2}-(|\psi _{2}|^{2}+\gamma |\psi _{1}|^{2})\psi _{2}-\lambda \left( \frac{%
\partial \psi _{1}}{\partial x}+i\frac{\partial \psi _{1}}{\partial y}%
\right) ,  \label{Ben2}
\end{equation}%
where $\nabla ^{2}$ is the Laplacian acting on coordinates $\left(
x,y\right) $, the real-valued $\lambda $ is the strength of the SOC, and $\gamma $ is
the relative strength of inter-component nonlinearity in comparison with the
intra-component self-attraction. Due to the specific form of the SOC terms,
composite solitons are generated by the system (\ref{Ben1})-(\ref{Ben2}) as
bound states of a fundamental localized state in one mode and a vortex in
the other mode (\textit{semi-vortices}, also referred to as filled vortices~\cite%
{bpafilled} or vortex-bright solitary waves~\cite{VB}), or mixtures of
fundamental and vortical components in both modes. As mentioned above,
solitons trapped in the OL potential are stabilized by it because their norm
drops below the threshold necessary for the onset of the collapse, while it
was believed that this is impossible in the 2D free space. The new feature
of the composite solitons produced by the system (\ref{Ben1})-(\ref{Ben2}) is
that their norm also takes values \emph{below the threshold}, without the
help of any trapping potential. The full stability is provided by the
comparison of values of the Hamiltonian of Eqs.~(\ref{Ben1})-(\ref{Ben2}),
\begin{eqnarray}
H&=&\int \int \left\{ \frac{1}{2}\left( |\nabla \psi _{1}|^{2}+|\nabla \psi
_{2}|^{2}\right) -\frac{1}{2}\left( |\psi _{1}|^{4}+|\psi _{2}|^{4}\right)
-\gamma |\psi _{1}|^{2}|\psi _{2}|^{2}\right.  \notag \\
&&\left.+\frac{\lambda }{2}\left[ \psi _{1}^{\ast }\left( \frac{\partial \psi
_{2}}{\partial x}-i\frac{\partial \psi _{2}}{\partial y}\right) +\psi
_{2}^{\ast }\left( -\frac{\partial \psi _{1}}{\partial x}-i\frac{\partial
\psi _{1}}{\partial y}\right) \right] +\mathrm{c.c.}\right\} dxdy,  \label{E}
\end{eqnarray}
where $\mathrm{c.c.}$, as well as the asterisk, stands for the complex
conjugate, for the composite semi-vortex and mixed-mode solitons. The
self-trapped modes of the former and latter types are stable and realize the
system's ground state (i.e., they minimize energy (\ref{E})) at $\gamma <1$
and $\gamma >1$, respectively.

\subsubsection{Repulsive BEC}

For the repulsive interaction between atoms, it has been predicted that
stable matter-wave vortices are supported by condensates loaded into OLs
\cite{MKO} and that gap solitons and gap-soliton vortices also exist if the
condensate is loaded into the OL of the same dimension \cite{15}. On the
other hand, the application of an OL, or of a magnetic lattice \cite%
{magn-latt}, to impose spatial modulation of the scattering length, $%
a_{s}\left( x,y,z\right) $ in Eq. (\ref{GPE}), induces an effective
nonlinear lattice, which can readily support 1D solitons, while the
stabilization of their 2D counterparts by nonlinear periodic potentials is a
difficult problem \cite{RMP}.

New perspectives for the creation of stable complex 3D localized modes, such
as vortex rings, vortex-antivortex hybrids, and \textit{Hopfions} (twisted
rings, featuring two independent topological numbers), which were
unavailable in other physical media, were recently predicted by a model with
the local strength of the self-repulsive cubic nonlinearity growing from the
center to periphery at any rate faster than $r^{3}$ \cite%
{Barcelona0,gyroscope,Barcelona}. Realization of these settings in BEC is a
challenge to the experiment. Below we will focus on some of
the more standard settings involving vortices and related structures in
higher-dimensional BECs without spatial or temporal modulation of the
scattering length.

\subsection{Vortices and vortex clusters}

Arguably, one of the most striking features of BECs is the possibility of
supporting vortices, which have been observed
in many experiments by means of a variety of methods. Vortices are
characterized by their non-zero topological charge $S$, whereby the phase of
the wavefunction has a phase jump of $2\pi S$ along a closed contour
surrounding the core of the vortex. The width of single-charge vortices in
BECs is of the order $\mathcal{O}(\xi)$ -- where $\xi$ is the healing length
of the condensate (see, e.g., \cite{PitStrin}) -- while higher-charge
vortices, with $|S|>1$, have cores wider than the healing length. Such
higher-charge vortices are generally unstable in the homogeneous background
case; nevertheless, they may be stabilized by employing external impurities~%
\cite{simula:pra2002} (the latter can be used to confine the so-called
persistent current; see, e.g., the recent discussion of relevant experiments
in~\cite{bpapers} and references therein), or by using external potentials
\cite{pu99}. Notice that, when unstable, higher-charge vortices typically
split in multiple single-charge vortices, since the system has no other way
to dispose of the topological charge~\cite{ketterle_vort}.

The fact that single-charge vortices carry topological charge renders them
extremely robust objects: indeed, continuous deformation of the vortex
profile cannot eliminate the $2\pi$ phase jump. An exception is a case where
the background condensate density is close to zero, and that is why, in BEC
stirring experiments, vortices are nucleated at the periphery of the
harmonically trapped condensate \cite{Gardetal}.

Vortices are prone to motion caused by gradients in the density (and phase)
of the background, induced by an external potential (as, e.g., in the case
of a trapped BEC) or by the presence of other vortices. The motion of the
vortex in such cases can be studied by means of the matched asymptotic
expansion method \cite{rubi}. The same method can also be used to study the
effect of vortex precession induced by the external trap (see, e.g., the
review \cite{9}), also in the presence of collisional inhomogeneities and
dissipative perturbations \cite{ourjpb}. Note that in the simplest case of a
single vortex in a 2D BEC confined in a harmonic trap, a Bogoliubov-de
Gennes analysis reveals the connection of the vortex precession frequency
with characteristic eigenfrequencies of the spectrum and -- in particular --
with the \textit{anomalous mode}  \cite{fettam} (for the latter, the integral of the norm~$%
\times$~energy product is negative \cite{PitStrin}). Indeed,
the negative-energy mode bifurcates in the linear limit from the dipole mode
(which has a constant magnitude equal to the trap frequency); then, as the
chemical potential increases, the anomalous mode eigenfrequency decreases,
and becomes equal to the precession frequency of the vortex in the
Thomas-Fermi limit (see, e.g., Refs.~\cite{ourjpb,ourdsv}).

On the other hand, the motion induced on a vortex by another vortex is
tantamount to the one observed in fluid vortices (see, e.g. Ref.~\cite%
{pismen}): this way, vortices with same charge travel parallel to each other
at constant speed, while vortices of opposite charges rotate about each
other at constant angular speed. Following Helmholtz' and Kirchhoff's
considerations, one may treat vorticity as a sum of point vortices and
determine the velocity field created by the vortices (this velocity field
induces the vortex motion) by means of the Biot-Savart law. This way, one may
find a set of ordinary differential equations (ODEs) for the location of the
vortices, that describe vortex-vortex interactions \cite{newton}. A more
subtle consideration in this context is the ``screening'' effect of the
vortex-vortex interaction by the inhomogeneities in the density (due to the
presence of the external potential). The latter effect has been incorporated
in some of the above works by an effective renormalization of the
interaction prefactor, but a more systematic study of this effect is still
lacking despite the fact that the relevant more complicated dynamical
equations can still be systematically derived~\cite{busch_mc}.

In a 2D (disk-shaped) BEC confined in a parabolic trap, it is then possible
to employ variational arguments \cite{peli} and combine both effects: vortex
precession (induced by the trap) and vortex-vortex interactions. This way,
the effective dynamics of a small cluster of interacting vortices (of
potentially same or different charges) is described by a system of ODEs for
the centers of the vortices. The relevant dynamical system possesses two
integrals of motion (Hamiltonian and angular momentum) and, thus, it is
completely integrable for vortex dipoles composed by two counter-rotating or
co-rotating vortices. Importantly, a theoretical description of vortex
trajectories was found to be in excellent agreement with pertinent
experimental findings ~\cite{dshstuff3}. Notice that apart from vortex
dipoles, the same methodology has been used in cases of vortex clusters
composed by more than two vortices. Clusters involving e.g. $4$, $6$, $8$
vortices of alternating charges in polygonal form~\cite{ourdsv} or $4$, $5$,
$6$ vortices in a linear configuration are currently challenging to produce
experimentally (and are found to be dynamically unstable when possessing more
than $4$ vortices in a polygonal shape, or 3 or more vortices in a linear
configuration~\cite{annabarry}). On the other hand, producing such clusters
with a controllable number of vortices of the same charge is straightforward (see the second item in Ref. \cite{dshstuff3}). Please note that the system possesses
a number of intriguing symmetry-breaking bifurcations ~\cite{alexandra}, which
can be explained even in analytical form through a linear stability analysis~%
\cite{theo}.

Apart from single vortices and small vortex clusters, there has been much
interest in \textit{vortex lattices} in rapidly rotating condensates. Such
configurations consist of a large number of ordered lattices of vortices,
arranged in triangular configurations, the so-called Abrikosov lattices~\cite%
{Abri}. The first BEC experiments reported observation of vortex lattices
consisted of just a few ($<15$) vortices \cite{MIT1}, but later on it was
possible to nucleate experimentally and maintain vortex lattices with over $%
100$ vortices \cite{vlat2,vlat3}. Subsequent efforts enabled the observation
of intriguing phenomena associated with these vortex lattices, including
their collective (so-called Tkachenko) oscillations, as well as their
structural phase transitions either under multi-component interaction (a
transition from hexagonal to square lattice was observed experimentally in~%
\cite{vlat4}) or in the presence of external potentials (a similar
transition was theoretically reported in the presence of a square optical
lattice in~\cite{vlat5}). Lastly, it is relevant to mention here that such
multi-vortex configurations are at the heart of ongoing studies of phenomena
including vortex turbulence and more generally non-equilibrium dynamics of
atomic condensates~\cite{turb}.

Admittedly, the above discussion contains a rather partial perspective of a
field that has truly boomed in a remarkable number of research threads and
has done so via a rather unique cross-pollination with other fields of
physics that is simply impossible to capture within the confines of the
present chapter. Nevertheless, we hope to have conveyed some of the main
areas of the pertinent studies and the ever-expanding (in terms of research
groups and themes of study) enthusiasm surrounding this area. We now turn
our attention to the specific contributions associated with this special
volume, which touch upon many of the  above-mentioned topics.

\section{A synopsis of the articles included in the present Special Issue}

As mentioned above, the current theoretical and experimental work on BEC and
related topics covers a vast research area. Of course, the papers selected for
this Special Issue cannot survey all aspects of this work. Most of the
papers present theoretical results, in compliance with the obvious trend
that many more original theoretical papers on BEC, than experimental ones,
appear in the scientific literature. Nevertheless, some articles from the Special Issue
present experimental results too, as briefly recapitulated below. Some
papers deal directly with basic aspects of the studies of BEC, while others
address different but related topics, such as fermion quantum gases,
few-boson models, etc. The different settings and problems addressed in the
articles may be categorized as more physical or more mathematical ones.

(1.) M. A. Caracanhas, E. A. L. Henn, and V. S. Bagnato, \textit{Quantum
turbulence in trapped BEC: New perspectives for a long lasting problem}

BEC in atomic gases provides the most natural testbed for exploring
turbulent dynamics of superfluids. This article \cite{BEC20_1} offers a
review of recent experimental and theoretical results on this topic, and a
discussion on the directions for the further development of studies dedicated to
quantum-liquid turbulence.

(2.) A. Vardi, \textit{Chaos, ergodization, and thermalization with few-mode
Bose-Einstein condensates}

This article \cite{BEC20_2} considers a system with few degrees of freedom,
which represents ``small" Bose-Hubbard (BH) models, namely, BH dimers and
trimers, the former one being reduced to a classical kicked top. In the
framework of these systems, the analysis is focused on aspects of classical
dynamical chaos in them, including the problems of the onset of ergodicity
and thermalization. The energy diffusion in the systems' phase space is
explored by means of the Fokker-Planck equation.

(3.) R. Radha and P. S. Vinayagam, \textit{An analytical window into the
world of ultracold atoms}

The paper \cite{BEC20_3} addresses BEC models based on GPEs, in terms of the
possible integrability of these models. The approach develops the known
method of transforming the standard integrable form of the NLSE into
seemingly complex, but still integrable ones, by means of explicit
transformations of the wave functions and variables $\left( x,t\right) $. In
particular, considered are models which may be integrable, while
they include complex ingredients, such as the time dependence (management)
of the scattering length, and a parabolic potential (expulsive, i.e., of the
anti-harmonic-oscillator type, rather than the trapping one), with a
constant or time-dependent strength. In addition to the single-component
models, two-component systems are considered too, with both nonlinear and
linear couplings between the components. A number of exact solutions are
found in such models, including bright and dark solitons.

(4.) A. I. Nicolin, M. C. Raportaru, and A. Bala\v{z}, \textit{Effective
low-dimensional polynomial equations for Bose-Einstein condensates}

The article \cite{BEC20_4} addresses the derivation and analysis of
effective equations with reduced (1D and 2D) dimensions for prolate and
oblate (cigar-shaped and pan\-cake-shaped) condensates, respectively. The
equations with the reduced dimensionality are derived from the full 3D GPE,
under the condition of strong confinement in the transverse direction(s).
The effective equations with polynomial nonlinearities are derived in this
context.

(5.) V. I. Yukalov and E. P. Yukalova, \textit{Statistical models of
nonequilibrium Bose gases}

The analysis in this article \cite{BEC20_5} addresses strongly perturbed
BEC, i.e., it goes far beyond the limits of the near-equilibrium mean-field
theory. In particular, this paper makes a contact with the considerations
presented in the article by M. A. Caracanhas, E. A. L. Henn, and V. S.
Bagnato in the same Special Issue \cite{BEC20_1}, as the analysis develops a
description of strongly excited BEC in terms of a statistical model of grain
turbulence.

(6.) H.-S. Tao, W. Wu, Y.-H. Chen, and W.-M. Liu, \textit{Quantum phase
transitions of cold atoms in honeycomb optical lattices}

Optical lattices with different geometries (in particular, 2D honeycomb
lattices considered in this article) help to support quantum gases (both
bosonic and fermionic) in highly-correlated states. The article \cite%
{BEC20_6} aims to review recent results for quantum phase transitions of
cold fermionic atoms in these lattices. In that sense, it is an essential
addition to the collection of topical chapters on the theme of BEC, as
results are presented for quantum Fermi gases, rather than for bosons. The
analysis combines mean-field considerations with quantum Monte-Carlo
computations, with the aim to calculate various properties of the systems
under consideration, such as the density of states, the Fermi surface, etc.
Also considered in this article are bilayer lattices, in addition to the
monolayer ones, and effects of the spin-orbit interaction between the
fermion components.

(7.) T. He, W. Li, L. Li, J. Liu, and Q. Niu, \textit{Stationary solutions
for nonlinear Schr\"{o}dinger equation with ring trap and their evolution
under the periodic kick force}

This work \cite{BEC20_7} addresses solutions of the nonlinear Schr\"{o}dinger equation for a
periodically kicked quantum rotator. The model may also be relevant for a
BEC trapped in a toroidal quasi-1D trap, with a periodically applied potential
profile. The analysis is focused on the especially interesting cases of
quantum anti-resonance and quantum resonance, using analytical stationary
solutions of the nonlinear Schr\"{o}dinger equation with periodic boundary conditions.

(8.) D. A. Zezyulin and V. V. Konotop, \textit{Stationary vortex flows and
macroscopic Zeno effect in Bose-Einstein condensates with localized
dissipation}

This article \cite{BEC20_8} addresses an interesting topic of nonlinear BEC
in dissipative media. A specific setting is considered with flow of the
superfluid towards the central part of the 2D system, where the loss is
concentrated. The model does not include any explicit gain, but,
nevertheless, it gives rise to stationary global patterns, including those
with embedded vorticity, due to the balance between the influx from the
reservoir at infinity and the effectively localized dissipation. The
solution is interpreted in terms of the Zeno effect in the dissipative BEC.

(9.) V.$\;$ Achilleos,$\;$ D.$\;$ J.$\;$ Frantzeskakis,$\;$ P.$\;$ G.$\;$ Kevrekidis,$\;$ P.$\;$ Schmelcher,$\;$ and J. Stockhofe, \textit{Positive and negative mass solitons in spin-orbit
coupled Bose-Einstein condensates}

The article \cite{BEC20_9} addresses the currently hot topic of solitons in
the two-com\-po\-nent BEC realizing the spin-orbit-coupling effect. The
analysis is developed for the 1D geometry, and relies on the reduction of
the underlying two-component GPE system to a single NLSE, by means of the
multiscale-expansion method. In this way, apart from the usual positive-mass
bright and dark solitons, negative mass structures, namely bright (dark)
solitons for repulsive (attractive) interactions are predicted as well. The
analytical predictions are confirmed by numerical simulations.

(10.) A. I. Yakimenko, S. I. Vilchinskii, Y. M. Bidasyuk, Y. I.
Kuriatnikov, K. O. Isaieva, and M. Weyrauch, \textit{Generation and decay of
persistent currents in a toroidal Bose-Einstein condensate}

The topic of persistent superfluid flows in toroidal traps is theoretically
addressed in the article, being motivated by recent experimental
observations of this effect. This article \cite{BEC20_10} offers a review of
theoretical results on this topic, recently produced by the present authors.
In particular, special attention is paid to the phenomenon of hysteresis in
this setting. The analysis is performed by means of numerical solutions of
the 3D GPE. Some related 2D settings are considered too.

(11.) M. Galante, G. Mazzarella, and L. Salasnich, \textit{Analytical
results on quantum correlations of few bosons in a double-well trap}

This work \cite{BEC20_11} deals not with BEC proper, but rather with sets of
few bosons, the number of which is $N=2,$ $3$, or $4.$ The bosons are
trapped in a double-well potential, which is also used in many experimental
and theoretical studies of BEC, such as those dealing with bosonic Josephson
oscillations. Eventually, the system is reduced to the simplest two-site
truncation of the Bose-Hubbard model, which has something in common with the
setting considered in another article included into this Special Issue, the
one by Vardi \cite{BEC20_2}. The analysis aims to find exact ground states for these
few-boson sets, and study variation of their characteristics, such as the
energy and entanglement entropy, as functions of system's parameters.

(12.) V. Bolpasi and W. von Klitzing, \textit{Adiabatic potentials and atom
lasers }

The article \cite{BEC20_12} addresses the topic of the design of atom-beam
lasers. A detailed analytic model of the trap is presented and the flux of the atom 
laser is determined. The analytical results are found to  be in good agreement
with recent experimental data. The analysis is focused on the harmonic-oscillator
trapping potential for the BEC, from which the laser beams are emitted.
Gravity is taken into regard too.

\end{document}